\documentclass[runningheads]{llncs}

\usepackage{color}
\usepackage{caption}
\usepackage{graphicx}
\usepackage{subcaption}
\usepackage{hyphenat}
\usepackage{url}
\usepackage{xspace,framed}

\makeatletter
\RequirePackage[bookmarks,unicode,colorlinks=true]{hyperref}%
   \def\@citecolor{blue}%
   \def\@urlcolor{blue}%
   \def\@linkcolor{blue}%

\def\orcidID#1{\smash{\href{http://orcid.org/#1}{\protect\raisebox{-1.25pt}{\protect\includegraphics{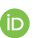}}}}}
\makeatother

\begin{document}

\title{\tool: A White-Box Fuzzer for Finding Security Vulnerabilities in C Programs}
\titlerunning{A White-Box Fuzzer for Finding Security Vulnerabilities in C Programs}

\author{Kaled M. Alshmrany
\inst{1}\orcidID{0000-0002-5822-5435} \thanks{Jury Member} \and
        Rafael S. Menezes\inst{2}\orcidID{0000-0002-6102-4343} \and
        Mikhail R. Gadelha\inst{3}\orcidID{0000-0001-6540-6587} \and
        Lucas C. Cordeiro\inst{4}\orcidID{0000-0002-6235-4272}}
\authorrunning{K. M. Alshmrany et al.}
\institute{
    University of Manchester, Manchester, UK\\ Institute of Public Administration, Jeddah, Saudi Arabia\\ \email{kaled.alshmrany@manchester.ac.uk} \and
  Federal University of Amazonas, Manaus, Brazil\and
    SIDIA Instituto de Ci\^encia e Tecnologia, Manaus, Brazil\and
  University of Manchester, Manchester, UK\\ 
}

\newcommand\tool{{\sf FuSeBMC}\xspace}

\maketitle

\begin{abstract}
We describe and evaluate a novel white-box fuzzer for C programs named \tool, which combines fuzzing and symbolic execution, and applies Bounded Model Checking (BMC) to find security vulnerabilities in C programs.
\tool explores and analyzes C programs (1) to find execution paths that lead to property violations and (2) to incrementally inject labels to guide the fuzzer and the BMC engine to produce test-cases for code coverage.
\tool successfully participates in Test-Comp'21 and achieves first place in the \texttt{Cover-Error} category and second place in the \texttt{Overall} category.
\end{abstract}

\section{Test Generation Approach}
\label{sex:overview}

\begin{figure}
    \centering
    \includegraphics[width=\textwidth]{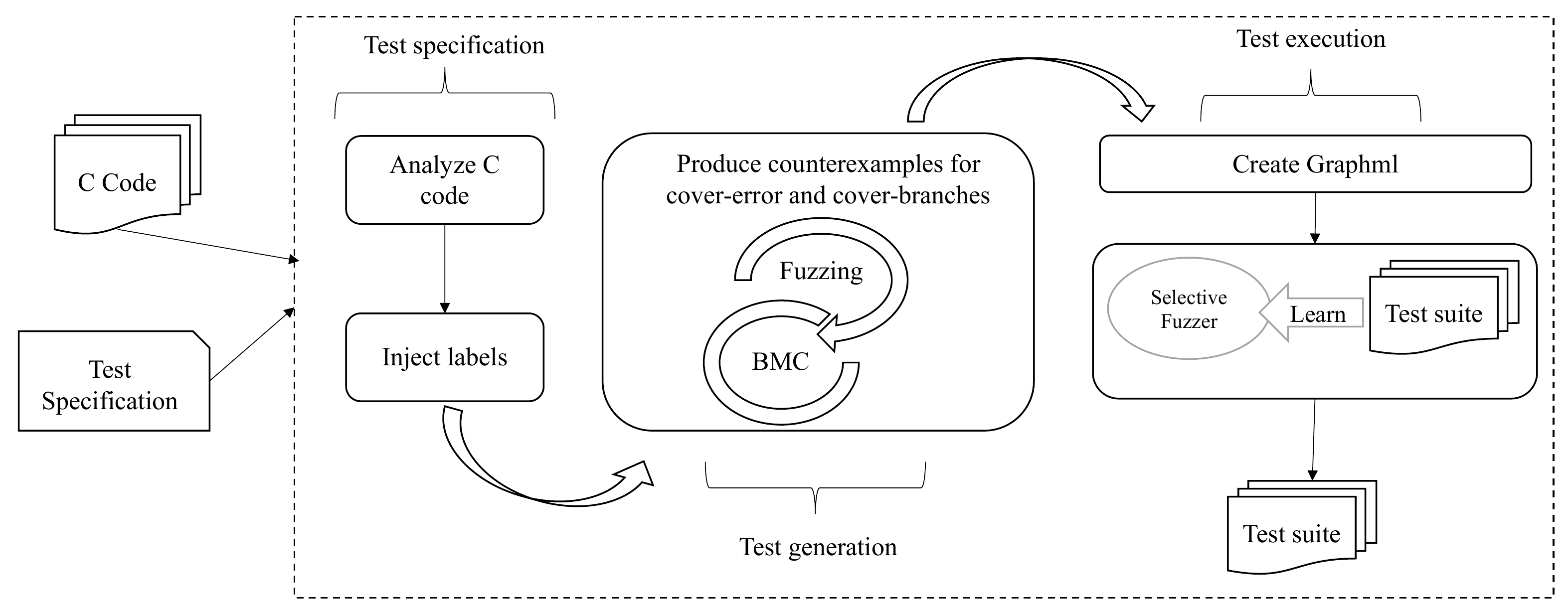}
    \caption{\tool: a white-box fuzzer framework for C Programs.}
    \label{fig:framework}
\end{figure}

Automated test-case generation is a method to check whether the software matches expected requirements~\cite{AnandBCCCGHHMOE13}. It involves the automated execution of software components to evaluate intricate properties and achieve code coverage metrics (e.g., decision, branch, instruction). Here, we describe and evaluate a novel white-box fuzzer, \tool, capable of automatically producing test-cases for C  programs.
\tool provides an innovative software testing framework that detects security vulnerabilities in C programs by using fuzzing and symbolic execution in combination with Bounded Model Checking (BMC) (cf. Fig.~\ref{fig:framework}). \tool builds on top of clang~\cite{CLANG} to instrument the C program, uses Map2check~\cite{menezes2018map2check} as a fuzzing engine, and ESBMC (Efficient SMT-based Bounded Model Checker)~\cite{esbmc2018} as a BMC engine, thus combining dynamic and static verification techniques.

\tool takes a C program and a test specification~\cite{TESTCOMP20} as input. In the \texttt{Cover-Error} category, \tool invokes the fuzzing and BMC engines to find a path that violates a given property. It uses an iterative BMC approach that incrementally unwinds the program until it finds a property violation or exhausts time or memory limits. \tool uses incremental BMC to explore the program state space searching for a property violation since all programs in Test-Comp'21 are known to have issues. In the \texttt{Cover-Branches} category, \tool explores and analyzes the target C program using the clang compiler to inject labels incrementally. \tool will compute all branches and inject the labels for each branch by adding the label \texttt{GOAL-$N$}, where $N$ is the goal number. Both engines will check whether these injected labels are reachable to produce test-cases for branch coverage.

\tool analyzes the counterexamples and saves them as a \textit{graphml} file. It checks whether the fuzzing and BMC engines could produce counterexamples for both categories \texttt{Cover-Error} and \texttt{Cover-Branches}. If that is not the case, \tool employs a second fuzzing engine based on selective fuzzer to produce test-cases for the rest of the labels. The selective fuzzer produces test-cases by learning from the two engines' output: it analyzes the range of the inputs that should be passed to examine the target C program and then produce different test-cases. Lastly, \tool prepares valid test-cases with metadata to test a target C program using TestCov~\cite{TESTCOMP20} as a test validator.

\tool sets a $150$ seconds limit for the fuzzing engine and a $700$ seconds limit for the BMC engine and sets a $50$ seconds limit for the selective fuzzer. These numbers were obtained empirically by analyzing the Test-Comp'21 results.

\section{Strengths and Weaknesses}
\label{sec:strengths-weaknesses}

Incremental BMC allows \tool to keep unwinding the program until a property violation is found or time or memory limits are exhausted. This approach is advantageous in the \texttt{Cover-Error} category as finding one error is the primary goal. Another strength of \tool is that it can accurately model C programs that use the IEEE floating-point arithmetic~\cite{GadelhaMMCN20}. The floating-point encoding layer in our BMC engine extends the support for the SMT \texttt{FP} theory to solvers that do not support it natively. \tool can test programs with floating-point arithmetic using all currently supported solvers in  BMC engine (ESBMC), including Boolector~\cite{NiemetzPreinerBiere-JSAT15}, which does not support the SMT \texttt{FP} theory natively.

In both \texttt{Cover-Error} and \texttt{Cover-Branches} categories, various test-cases produced by \tool are validated successfully. The majority of our test-cases were produced by the BMC engine and the selective fuzzer; our fuzzing engine did not produce many test-cases because it does not model the C library, so it mostly guesses the inputs. However, our fuzzing engine is not limited to only produce test-cases: it helps our selective fuzzer by providing information about the number of inputs required to trigger a property violation, i.e., the number of assignments required to reach an error. In several cases, the BMC engine can exhaust the time limit before providing such information, e.g., when there are large arrays that need to be initialized at the beginning of the program.

Apart from that, our employed verification engines also demonstrate a certain level of weakness to produce test-cases due to the many optimizations we perform when converting the program to SMT. In particular, two techniques affected the test-case generation significantly: \textit{constant folding} and \textit{slicing}. \textit{Constant folding} evaluates constants (which includes nondeterministic symbols) and propagates then throughout the formula during encoding, and \textit{slicing} removes expression not in the path to trigger a property violation. These two techniques can significantly reduce SMT solving time. However, they can remove the expressions required to trigger a violation when the program is compiled, i.e., variable initialization might be optimized away, forcing \tool to generate a test-case with undefined behavior.

Regarding our fuzzing engine, we identified a limitation to handle programs with pointer dereferences. The fuzzing engine keeps track of variables throughout the program but has issues identifying when they go out of scope. When we try to generate a test-case that triggers a pointer dereference, our fuzzing engine provides thrash values, and the selective fuzzer might create test-cases that do not reach the error.

\section{Tool Setup and Configuration}
\label{sec:setup}

In order to run our \texttt{fusebmc.py} script\footnote{\url{https://gitlab.com/sosy-lab/test-comp/archives-2021/-/blob/master/2021/FuSeBMC.zip}}, one must set the architecture ({\it i.e.}, $32$ or $64$-bit), the competition strategy (i.e., \textit{k}-induction, falsification, or incremental BMC), the property file path, and the benchmark path, as:

\begin{verbatim}
fusebmc.py [-a {32, 64}] [-p PROPERTY_FILE]
                 [-s {kinduction,falsi,incr,fixed}]
                 [BENCHMARK_PATH]
\end{verbatim}

\noindent where \texttt{-a} sets the architecture, \texttt{-p} sets the property
file path, and \texttt{-s} sets the strategy (e.g., {\tt kinduction}, {\tt falsi}, {\tt incr}, or {\tt fixed}). For Test-Comp'21, \tool uses \texttt{incr} for incremental BMC.

By choosing fuzzing engine, we set the following options when executing Map2Check: timeout of 150 seconds for Map2Check in \texttt{Cover-Error}, and a timeout of 70 seconds in \texttt{Cover-Branches};
\texttt{--fuzzer-mb $1000$} limits memory to $1000$ MB;
\texttt{--target-function-name reach$-$error} defines the function name to be searched;
\texttt{--target-function} checks whether the target-function is reachable;
\texttt{--nondet-generator fuzzer} uses only libfuzzer~\cite{serebryany2015libfuzzer};
\texttt{--generate-witness} sets the witness output path.

By choosing incremental BMC, the following options are set when executing ESBMC:
\texttt{--no-div-by-zero-check} disables the division by zero check (required by Test-Comp);
\texttt{--force-malloc-success} sets that all dynamic allocations succeed (a Test-Comp requirement);
\texttt{--floatbv} enables floating-point SMT encoding;
\texttt{--incremental-bmc} enables incremental BMC;
\texttt{--unlimited-k-steps} removes the upper limit of iteration steps for incremental BMC;
\texttt{--witness-output} sets the witness output path;
\texttt{--no-bounds-check} and \texttt{--no-pointer-check} disable bounds-check and pointer-safety checks, resp., since we are only interested in finding reachability bugs;
\texttt{--k-step 5} sets the incremental BMC to $5$;
\texttt{--no-allign-check} disables pointer alignment checks; and
\texttt{--no-slice} disables slicing of unnecessary instructions.

The Benchexec tool info module is named \texttt{fusebmc.py} and the benchmark definition file is \texttt{FuSeBMC.xml}.

\section{Software Project}
\label{sec:project}

The \tool source code is written in C++ and it is available for downloading at GitHub\footnote{\url{https://github.com/kaled-alshmrany/FuSeBMC}}, which includes the latest release of  \tool v3.6.6.
\tool is publicly available under the terms of the MIT License.
Instructions for building \tool from the source code are given in the file \texttt{README.md} (including the description of all dependencies).
%







\end{document}